\documentclass[10pt,letterpaper]{article}
\usepackage{graphicx}
\usepackage{amsmath}
\usepackage{cite}
\graphicspath{{./figures/}}

\RequirePackage[letterpaper]{geometry}
\RequirePackage{color}

\usepackage{mathptmx,courier,textcomp}
\usepackage[scaled=.92]{helvet}

\newcommand\OEtitle[1]{\LARGE \bf \hskip2.25pc \parbox{.8\textwidth}{ \noindent%
   \LARGE \bf \begin{center} #1 \end{center}\rm } \vskip.1in \rm\normalsize }

\newcommand\OEauthor[1]{\hskip2.25pc \parbox{.8\textwidth}{ \noindent%
   \normalsize \bf \begin{center} #1 \end{center}\rm } \vskip-1pc }

\let\title\OEtitle
\let\author\OEauthor

\let\address\OEaddress
\let\email\OEemail

{\vskip1pc\noindent\begin{center} \begin{minipage}{.8\textwidth} {\bf Abstract: \ } }
{ \\ \vskip-.5pc \noindent \small \copyright \, \number\year \hskip.05in
   Optical Society of America \\ \hfil \end{minipage}\end{center}\normalsize\vskip-1.5pc}%

\begin{document}


\title{Frequency stability of a dual-mode whispering gallery mode optical reference cavity}

\author{L.M. Baumgartel,$^{1,2}$ R.J. Thompson,$^1$ and N. Yu$^{1,*}$}

\address{$^1$Jet Propulsion Laboratory, California Institute of Technology, 4800 Oak Grove Dr, \\Pasadena 91109, California, USA\\
$^2$Department of Physics and Astronomy, University of Southern California\\Los Angeles 90089, California, USA}

\email{nan.yu@jpl.nasa.gov} 


\section*{Abstract}
We report an investigation of laser frequency stabilization using a whispering gallery mode resonator that is temperature stabilized by a dual-mode technique. This dual-mode technique has yielded mode volume temperature instabilities at the nK level, suggesting that high frequency stability may also be reached. Here, we experimentally and theoretically investigate the dynamics of such a system and the important factors affecting the achievable frequency stability. We calculate that the dual-mode technique can reduce the effective fractional temperature coefficient of the reference system to 3.6$\times10^{-8}$~K$^{-1}$ within the temperature feedback bandwidth. We demonstrate a 1560 nm laser stabilized to 1.3$\times10^{-12}$ at 1~s and 1.1$\times10^{-10}$ at 1000~s, corresponding to a long-term drift of 21 kHz/hr.

\section{Introduction}
Whispering gallery mode resonators (WGMRs) demonstrate extremely high quality factor and correspondingly narrow linewidth optical resonances over a large wavelength range, making them attractive for laser frequency stabilization applications. In fluoride resonators, quality factor ($Q$) in excess of one billion is regularly demonstrated, and $Q>$6$\times 10^{10}$ has been achieved~\cite{grudinin_pra2006}, corresponding to a linewidth of $\approx$4.5~kHz -- comparable to that of high-finesse Fabry P\'erot (FP) cavities. With special annealing techniques, an even higher $Q$ of 3$\times 10^{11}$ has been demonstrated~\cite{savchenkov_opex2007}. The wide transparency window of a crystalline WGMR mitigates the need for expensive mirror-coating runs and facilitates use of a single cavity at multiple wavelengths, while their compact size and robustness to mechanical vibrations makes them well suited for applications requiring portable high-performance lasers. 

Impressive gains to laser linewidths and short-term stability have already been achieved using WGMRs. Linewidths of 13~kHz were demonstrated with a resonator acting as a transmission filter in a fiber laser cavity~\cite{sprenger_optlett2010}, and a 200~Hz linewidth was achieved through injection locking to a WGMR~\cite{liang_optlett2010}. When employed as an external frequency reference cavity, crystalline resonators have delivered Allan deviations of 6$\times 10^{-14}$ at 100~ms~\cite{alnis_pra2011}. 

Frequency stability of WGMR reference cavities is limited by the fact that, in contrast to an FP resonator, the light travels in a solid medium. At short time scales, thermorefractive noise defines the limit and has been analyzed theoretically~\cite{savchenkov_josab2007,chijioke_pra2012} and measured experimentally~\cite{gorodetsky_josab2004}. At longer time scales, stability is limited by temperature changes in and around the resonator. These changes pull the frequency both through thermorefraction and thermal expansion, though the latter dominates in the common resonator host crystals magnesium fluoride and calcium fluoride. Ultra-low expansion (ULE) glasses have thermal expansion coefficients about three orders of magnitude lower than common optical crystals, but their poor transmission qualities preclude their use as a WGMR material. 

Recently, dual-mode temperature stabilization schemes have been investigated as a means to address this heavy dependence of frequency upon temperature~\cite{strekalov_optexpr2011,baumgartel_cleo2012,fescenko_opex2012,delhaye_cleo2012}. In such schemes, the frequency spacing between modes with different \emph{temperature coefficients (tempcos)} of frequency provides an extremely sensitive measurement of the mode volume temperature and facilitates active stabilization thereof. Mode volume temperature instabilities less than 10~nK have been demonstrated~\cite{strekalov_optexpr2011,baumgartel_cleo2012}. But the WGMR represents a complex thermal system with several different sources, sinks and the resulting gradients. We find that a highly stabilized mode volume temperature does not guarantee frequency stability, a finding also suggested by other recent work~\cite{fescenko_opex2012}. 

Here, we investigate how the relative strengths of various heat sources affect the temperature distribution in the resonator. By performing numerical modeling, and measuring the locked laser's frequency change as a function of various thermal parameters, we gain an understanding of the mechanisms limiting performance and the steps necessary for their mitigation. In particular, we find that the dual-mode technique leads to unstable temperature gradients in the resonator that pull the frequency through thermal strain. Despite these limitations, the technique reduces the effective temperature coefficient of the WGMR reference cavity by more than an order-of-magnitude -- from 8.88$\times 10^{-6}$~K$^{-1}$ to 1.33$\times 10^{-7}$~K$^{-1}$. We achieve an optical frequency instability of 1.29$\times10^{-12}$ at one second, and a long-term drift of 21~kHz/hr.

\section{Dual-mode stabilization}
As described in detail previously~\cite{strekalov_optexpr2011}, the thermorefractive coefficients of orthogonally polarized modes in a birefringent resonator are significantly different. Therefore, the frequency spacing between ordinary and extraordinary modes depends heavily upon temperature. The dependence is given by $d(\Delta f)/dT=-f_0(\alpha_n^{(o)}-\alpha_n^{(e)})$, where $\alpha_n^{(o,e)}$ are the thermorefractive coefficients of the ordinary and extraordinary polarizations, and $f_0$ is the optical carrier frequency. For magnesium fluoride (MgF$_2$) at our experimental temperature of 40~$^{\circ}$C and $f_0=c/$1560 nm, the differential tempco is 67~MHz/K. With a multi-mode resonator such as this, one can easily identify an appropriate pair of modes with spacing in the tens-of-megahertz range because, for each polarization, there are several tens of high-$Q$ modes within the free spectral range of 10.6~GHz. With the narrow linewidths in a MgF$_2$ resonator, a frequency spacing measurement on the Hz level is readily achieved, yielding nK sensitivity.  

\begin{figure}[ht!]
\centering
\includegraphics[width=13cm]{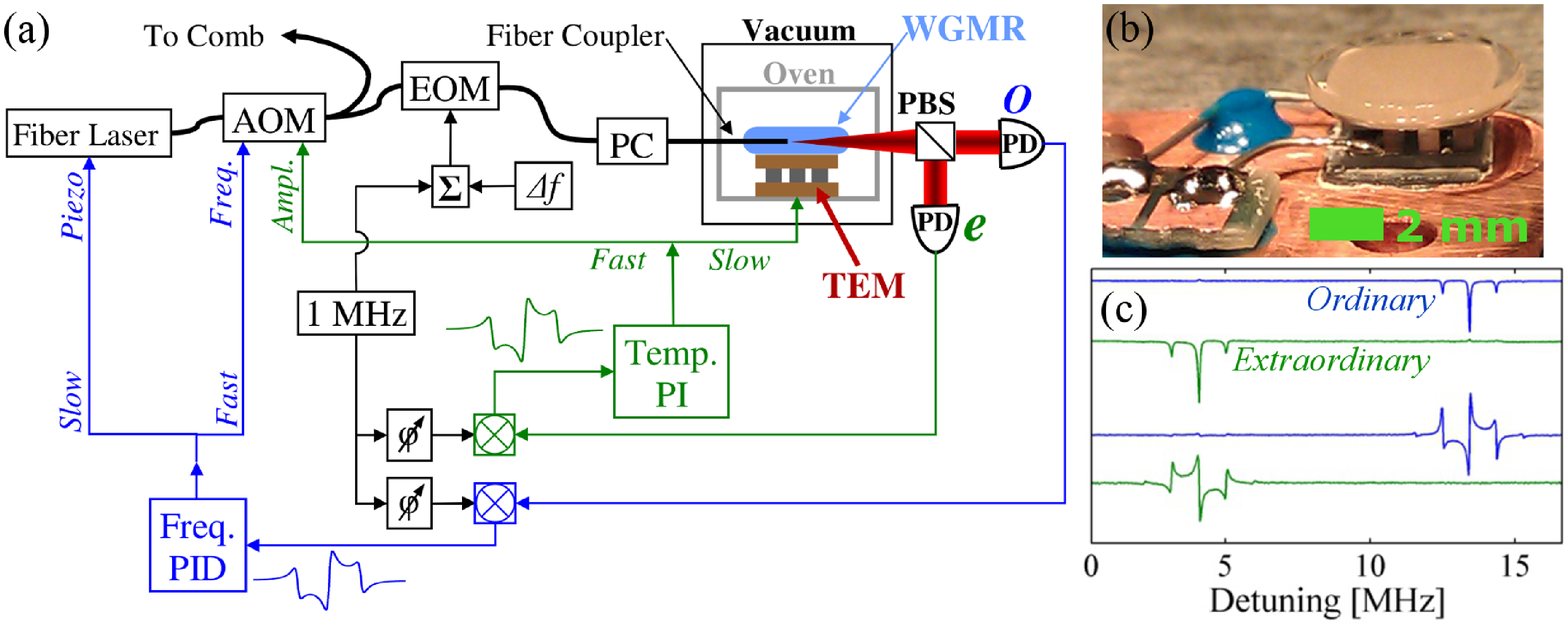}
\caption{The experimental setup. (a) Schematic of the optical system and stabilization loops. Each proportional-integrator (PI) loop has both a slow and fast actuation branch. Polarization controller (PC) allows for control of excitation ratio between the two modes. (b) Photo of the resonator-TEM assembly. (c) Scope traces of the detector and PDH error signals as the laser is swept over the orthogonal mode pair; modulation frequency is 1~MHz.}
\label{experiment_2c}
\end{figure} 

\subsection{Experimental setup}
Our implementation of the dual-mode stabilization scheme is shown in Fig.~\ref{experiment_2c} and described as follows. A narrow linewidth fiber laser is coupled via an angle-polished fiber~\cite{ilchenko_optlett1999} into 3.25~mm radius, z-cut resonator made from excimer-grade MgF$_2$. Mode linewidths are in the range of 40-80 kHz, corresponding to quality factors of 2.4-4.8$\times$10$^9$, depending on the mode order, polarization, and loading conditions. Laser light passes first through an acousto-optic modulator (AOM), and then an electro-optic phase modulator (EOM), before being coupled into the cavity at a polarization angle that can excite both modes. Two modulation frequencies are applied at the EOM: a 1~MHz signal for generation of Pound-Drever-Hall error signals~\cite{drever_japb1983}, and a second signal at the frequency spacing $\Delta f$ between a chosen pair of orthogonal modes. Thus, the laser is frequency locked to the ordinary mode, while the modulation sideband coincides with the extraordinary mode, providing an error signal for temperature stabilization. For these experiments, the error signal has a noise-limited temperature discriminant of 140 nK~Hz$^{-1/2}$. Thermal feedback (unity gain bandwidth $\approx$1~Hz) to the resonator is provided by a $4\times 4$~mm$^2$ thermoelectric module (TEM) glued directly to the disc; an optional fast thermal loop actuates on the in-coupled laser intensity via the AOM. The resonator-TEM assembly is mounted in a small ($2.5\times 2.5\times 2.5$~cm$^3$) brass “oven,” whose temperature is measured with a thermistor and stabilized by feeding back to a resistive heater. 

The stability of the resonator coupling is critical to the frequency stabilization and its practical use, so we paid special attention to the fiber coupler mounting scheme. The small oven in which the resonator is mounted has a base that is machined from a single block of brass. It has a flexure arm where the fiber coupler is attached. Initial alignment of the fiber coupler is performed with a 3D piezo stage, after which the fiber is glued to the arm. A micro screw is used on the flexure arm for subsequent small gap adjustments. This mounting scheme provides stable operation for periods as long as months without any need for adjustment and with no noticeable change in the coupling condition.

The oven-resonator-coupler setup is placed in a cylindrical vacuum chamber that is evacuated to the milli-torr level. Temperature of the vacuum canister itself is not actively controlled. Use of the vacuum chamber eliminates convective heat transfer and provides isolation from changes in atmospheric pressure and humidity.

We measure and characterize the system by examining the in-loop error signals and by
measuring the beatnote between the locked laser and a self-referenced frequency comb. The comb is stabilized to a hydrogen maser that has a fractional instability less than 1$\times 10^{-13}~\tau^{-1/2}$ from 1 to 10,000~s.

\subsection{Simulation}
The dual-mode stabilized resonator system has complex thermal properties and dynamics. To understand the various factors influencing mode volume temperature, as well as the overall thermal distributions in the resonator, finite element method (FEM) modeling proves useful. Two-dimensional, axially symmetric modeling of the resonator-TEM assembly is performed using a commercial FEM package (COMSOL). Material properties are taken from the manufacturer datasheets or material property databases; model geometries are based as closely as possible on the physical system. Screenshots of the FEM model are shown in Fig.~\ref{sim}.

The objective is to determine the frequency stability of a WGMR whose mode volume temperature is stabilized to the nK level via the dual-mode technique. Our model accounts for the following heat transfer mechanisms: radiative coupling between resonator and oven, optical heating of the mode volume, heat pumped into/out of the resonator by the thermoelectric effect, and thermal conduction through the TEM. (We assume convective transfer is negligible in the vacuum chamber.) Here, we are primarily concerned with dynamics within the TEM feedback bandwidth, i.e., those with times-scales $>$1~s. We calculate the thermally-induced strains on the resonator's radius $r$ and how they pull the frequency through $\Delta f/f_0=-\Delta r/r$. The model does not account for thermorefraction because the mode volume temperature is taken to be stable, as indicated by the residual error signal of the temperature stabilization loop.

Radiative heat transfer between one body completely enveloped by another can be estimated as~\cite{schmidt_1966}
\begin{equation}
\label{q_rad}
q_{rad}=\frac{\sigma}{1/\epsilon_1+A_1/A_2(1/\epsilon_2-1)}\left(T_1^4-T_2^4\right),
\end{equation}
where $\epsilon$ is the emissivity, $A$ the area, $T$ the temperature, the subscripts indicate inner (1) and outer (2) bodies, and $\sigma$ is the Stefan-Boltzmann constant. For various values of temperature difference between resonator and enclosure, $q_{rad}$ is calculated using Eq.~\eqref{q_rad} and then applied as a boundary condition on the resonator surface. 

Laser light circulating in the mode is converted to heat through material absorption. For a laser locked on resonance (zero detuning), the heat generated by this absorption can be estimated through~\cite{carmon_optexpr2004} 
\begin{equation}
q_{laser}=I\eta \frac{Q}{Q_{abs}}
\label{q_laser}
\end{equation}
where $I$ is the input laser power, $\eta$ is the coupling efficiency, and $Q$ ($Q_{abs}$) are the observed (absorption limited) quality factors. This heat is included in the model as a volume heat source (units of J/m$^3$) applied to the mode volume (region shown in pink in Fig.~\ref{sim}(a)). The mode volume cross-sectional shape is an ellipse with major (minor) axes of 20 (10) $\mu$m, which approximates the distribution of the fundamental mode as calculated by a separate, full vectorial FEM simulation as described in~\cite{grudinin_josab2012}.

Heat generated in the TEM ($q_{TEM}$) is modeled as a volume source distributed over the top alumina plate. The bottom of the TEM is held at the enclosure temperature. Thus, we can model how the three heat sources $q_{rad}$, $q_{laser}$, and $q_{TEM}$ contribute to the mode volume temperature. An initial steady-state solution is found assuming reasonable values of in-coupled laser power, with $q_{TEM}=q_{rad}=0$, yielding values of disc radius and mode volume temperature. The dual mode condition is simulated by demanding that subsequent changes in $q_{rad}$ or $q_{laser}$ are compensated for by a change in $q_{TEM}$, maintaining the mode volume temperature at its initial value. The resulting strains (changes in radius) are converted to frequency, allowing us to model how a change in enclosure temperature or laser intensity that occurs while the dual-mode loops are closed will affect the frequency of the resonator. 

\section{Results}
The dual-mode scheme measures and stabilizes the temperature in the region occupied by the optical modes. Our loop guarantees that any uncontrolled (out-of-loop) change in thermal transport parameters will result in a compensating change in the level of thermal feedback actuation. As the relative strengths of these thermal parameters changes, the system moves along lines of constant mode volume temperature. These isotherms do not, however, represent lines of constant resonance frequency because the central portion of the resonator changes temperature and expands/contracts. The isotherm slope, i.e., the ratio of change in optical frequency to the change in a system parameter, represents an effective coefficient of frequency. 

\subsection{Environmental temperature dependence}
We simulated the dual-mode resonator's effective tempco by solving the FEM model for various temperature differences between resonator and enclosure ($T_1$ and $T_2$ in Eq. \eqref{q_rad}). The condition $T_1=T_2$ means the average resonator temperature is in equilibrium with the oven and represents the instant at which the dual-mode loops are closed. This $q_{rad}=0$ situation is solved first. Next, a small drift in $T_2$ is imposed, $q_{rad}$ is calculated, applied as a boundary condition on the resonator, and the necessary value of $q_{TEM}$ needed to return the mode-volume temperature to its initial value is found -- this represents the feedback action of the dual-mode loop. 

\begin{figure}[htb]
\begin{minipage}[b]{0.55\linewidth}
\centering
\includegraphics[width=7.4cm]{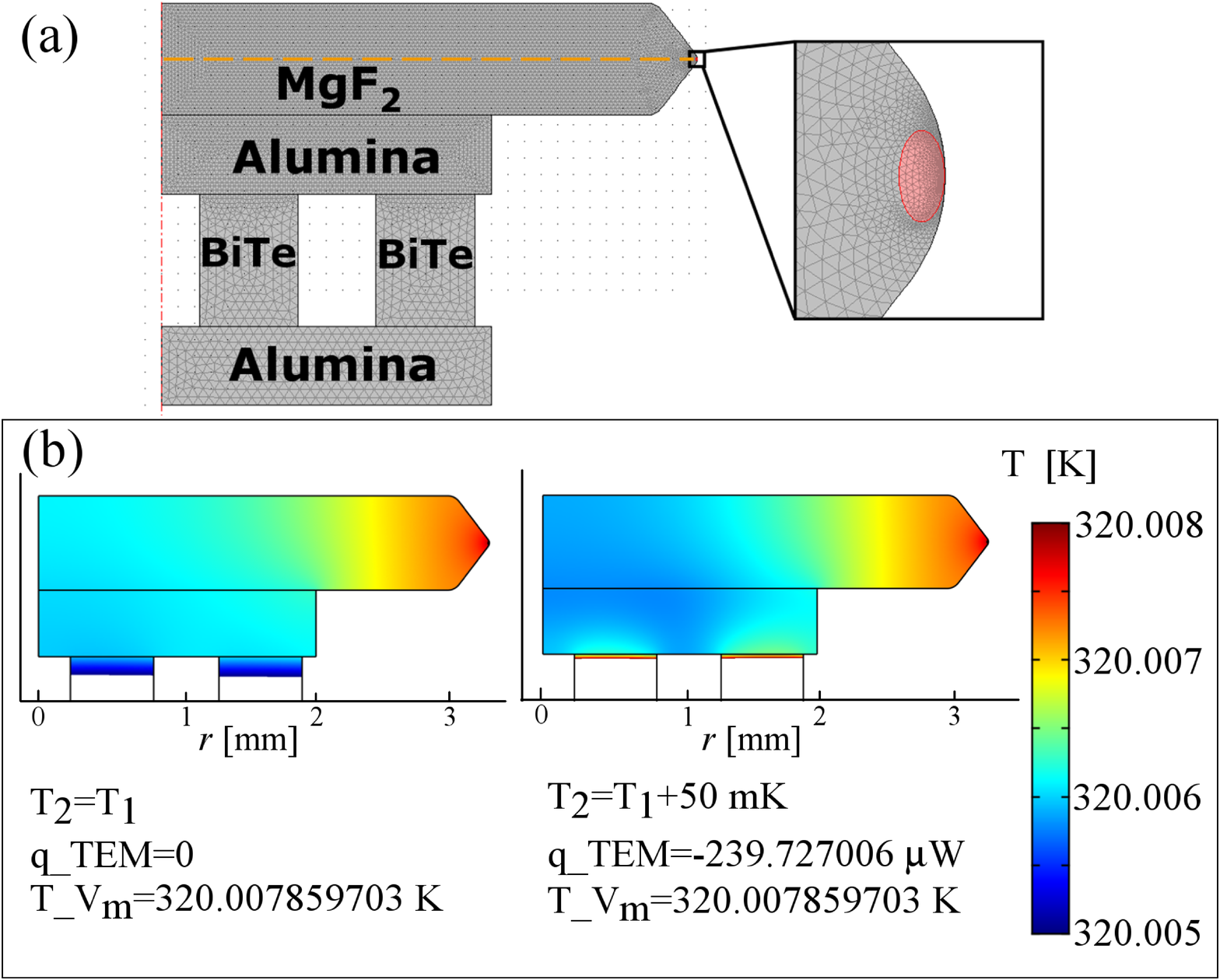}
\end{minipage}
\hspace{0.3cm}
\begin{minipage}[b]{0.4\linewidth}
\centering
\includegraphics[width=5.3cm]{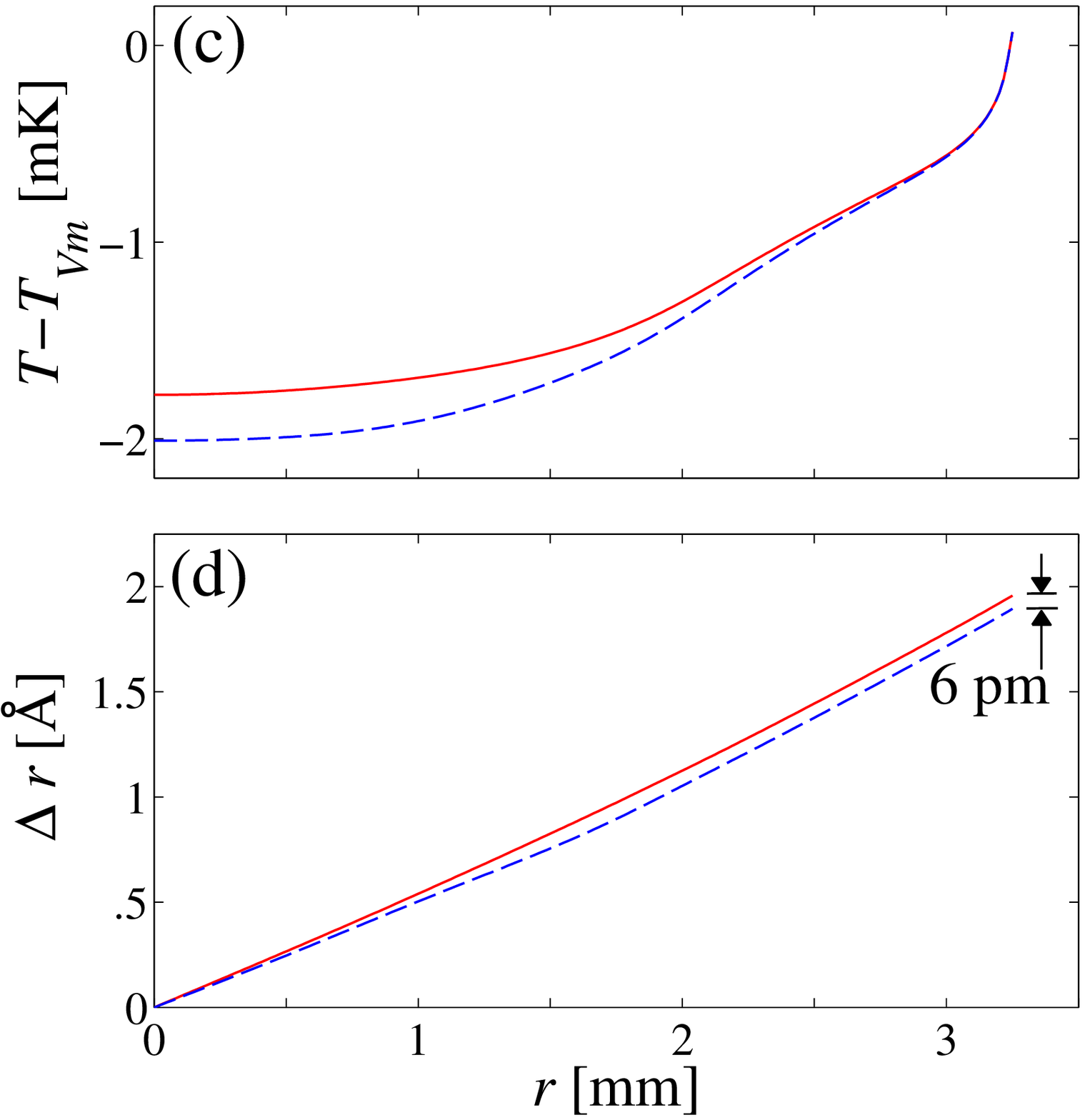}
\end{minipage}
\caption{FEM simulation. (a) Model setup showing the mesh, materials, a detail of the mode volume (pink region), and the line along which temperature and strain are plotted (dashed orange). BiTe: bismuth telluride. (b) Temperature map results showing how different gradients exist for different relative strengths of heat sources. Plots  of temperature (c) and radial deformation (d) along the disc radius for $T_2=T_1$ (solid red) and $T_2=T_1+50$~mK (dashed blue), showing how the latter distribution is cooler in the center, contracting the disc radius by $\approx$6~pm.}
\label{sim}
\end{figure}

The result is illustrated in Fig.~\ref{sim}(b), where the different thermal distribution resulting from a change in enclosure temperature is clearly visible. At $T_2=T_1$, no radiative heating of the disc occurs; the gradient forms as optically-generated heat is conducted out through the TEM. If the oven temperature drifts up by 50~mK, radiation additionally heats the top and outer portion of the disc. Maintaining the mode volume temperature -- out to the nK digit in these simulations -- requires that the TEM cool the central-bottom portion of the resonator, contracting it and increasing the common-mode optical frequency. The plots in Fig.~\ref{sim} show the calculated distribution of temperature (c) and deformation (d) along the dashed orange radial line in (a). The temperature distribution is normalized to the mode volume temperature ($T_{Vm}$), and the difference in strains at the endpoint ($R=3.25$~mm) yields the frequency shift. The calculated shift was found to be linear across five values of oven temperature drift spanning half a degree. A linear fit to these simulated data yields an effective tempco of 6.9~kHz/mK, corresponding to a fractional coefficient of 3.6~$\times 10^{-8}$~K$^{-1}$. We note that these values are comparable to the linear thermal expansion coefficients of common ULE glasses. 

\begin{figure}[htb]
\begin{minipage}[b]{0.5\linewidth}
\centering
\includegraphics[width=7cm]{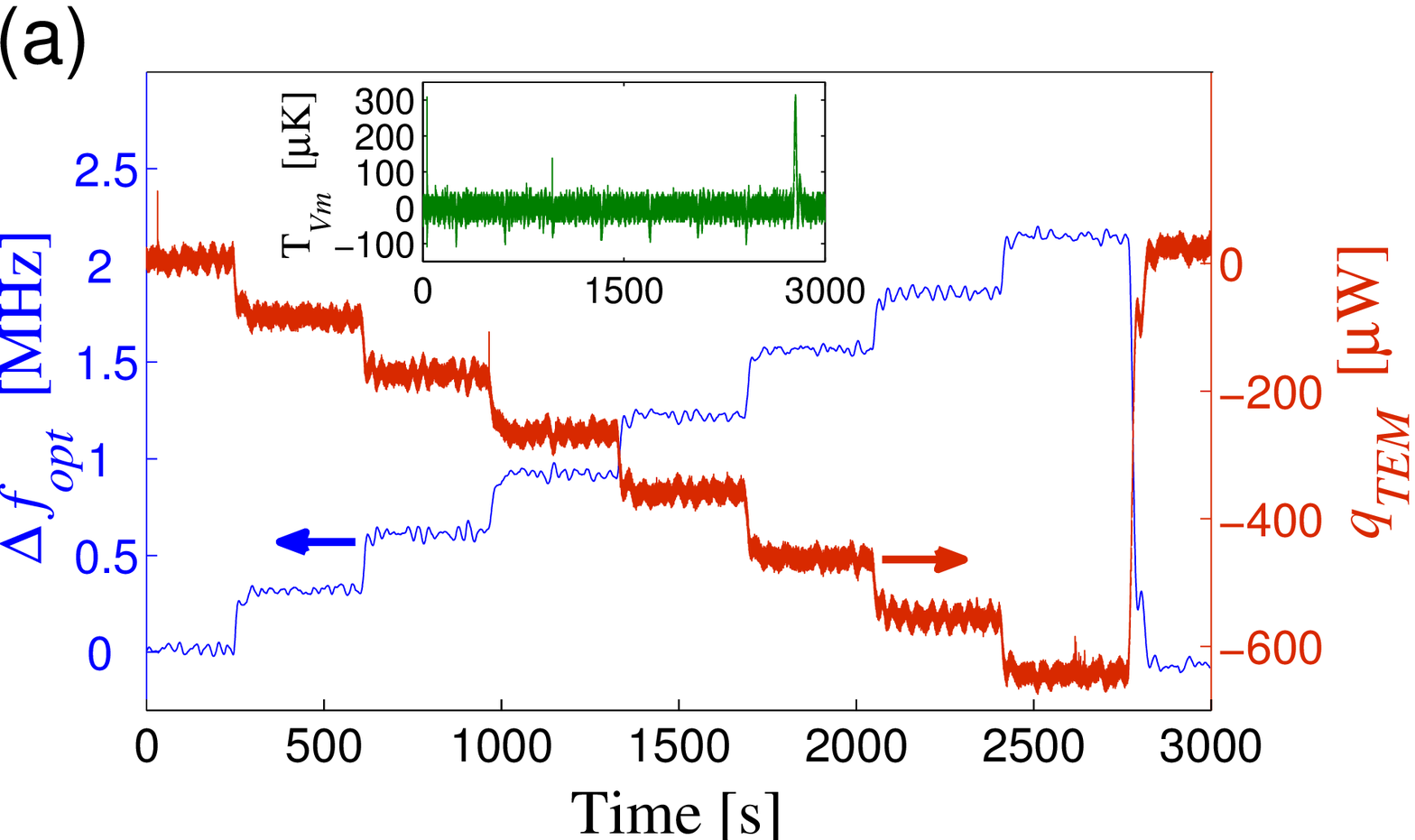}
\end{minipage}
\hspace{0.3cm}
\begin{minipage}[b]{0.47\linewidth}
\centering
\includegraphics[width=6.1cm]{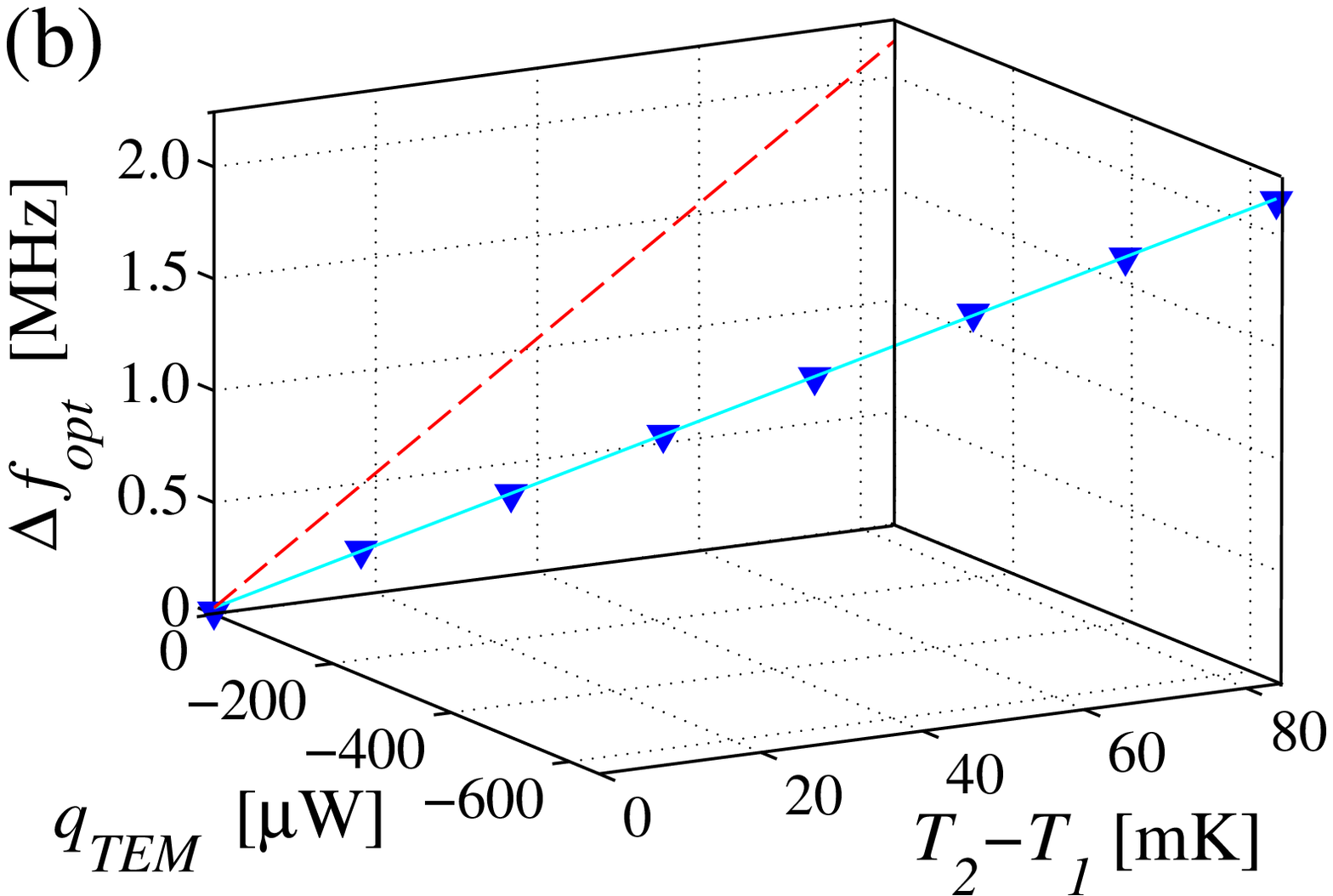}
\end{minipage}
\caption{Effect of environmental temperature: Time series (a) showing shift in optical frequency ($\Delta f_{opt}$) and heat delivered to the resonator by the TEM as enclosure temperature changes in 12~mK steps. Inset shows mode volume temperature from the in-loop error signal; aside from spikes at the temperature steps, it remains extremely stable. The multi-dimensional relationship between parameters is shown in (b). $T_2-T_1$ is the drift in enclosure temperature. Triangles are data from (a) averaged over 360 s bins. The isotherm is a fit to these data (solid blue line), while the slope of its projection onto the temperature-frequency axis (dashed red line) yields the effective frequency coefficient of 25.5~kHz/mK.}
\label{encl}
\end{figure}

The effective tempco was also measured experimentally. With the laser locked and the dual-mode stabilization loops engaged, we increased the setpoint on the oven's temperature controller in 12 mK steps every 360 seconds (eventually returning it to its initial value). The TEM current (converted to heat via values on the TEM's datasheet), locked laser's optical frequency, and in-loop temperature error signal were simultaneously recorded during the run, and are presented in Fig.~\ref{encl}. In agreement with the simulation, increased radiative heating of the disc results in compensatory cooling from the TEM, leaving the mode volume temperature the same but the optical frequency higher. A linear fit of the optical frequency as a function of change in enclosure temperature gives an effective tempco, with the experimental value being 25.5~kHz/mK. The magnitude of this value is 67 times less than the thermal-expansion determined tempco of -1.71~MHz/mK, and it corresponds to a fractional coefficient of 1.33$\times$10$^{-7}$~K$^{-1}$.

\subsection{Circulating power dependence}
Despite MgF$_2$'s extremely high transmission at 1560~nm, absorbed optical power is constantly converted to heat in the mode volume. Thus, changes in circulating optical power will destabilize the optical frequency at any time scale. Here, we investigate the effect of slow (within the TEM loop bandwidth) circulating power changes. These slower changes may be harder to control since they result from drifts in polarization or coupling strength, while faster fluctuations are in principle corrected for by the intensity branch of the dual-mode loop. Similar to the case of environmental temperature, we find a multi-parameter dynamic whereby a change in circulating power results in a change in TEM correction signal, thermal gradient, and resonator frequency.

Using the same FEM model described above, we calculated the frequency shift as a function of change in circulating optical power. For these simulations $q_{rad}$ was kept at zero. Heat was allowed to leave the resonator radiatively through its surface (to a boundary at infinity) and via conduction through the TEM. For a given change in circulating power, the resulting change in $q_{TEM}$ necessary to maintain the mode volume, change in thermal gradient, and shift in frequency was calculated. Using Eq.~\eqref{q_laser} and reasonable values of $I=100~\mu$W at the coupler, $\eta=0.07$, and that the disc's quality factor is absorption limited, the simulations suggest that a 5\% reduction in laser power will yield a frequency shift of -103~kHz.  

An experimental measurement of this isotherm slope was also made. With the laser intensity branch of the thermal loop open, we changed the optical power sent to the resonator by varying the amount of light transmitted through the AOM. In agreement with the simulations, a decrease in circulating power causes an increase in TEM heating to maintain the mode volume temperature. The new thermal gradient is warmer in the disc center, increasing $r$ and decreasing the optical frequency. The experimentally measured frequency pulling was -15~kHz for a 5\% reduction in intensity, corresponding to an effective coefficient of 43~kHz/$\mu$W of circulating optical power. Based on this result, a stability of $\sim$47~pW would be needed to reach a benchmark fractional frequency instability of $10^{-14}$. This requirement on laser intensity stabilization is a readily achieved 7~ppm, however we stress that the requirement is on circulating power, so coupling strength and polarization angle (which changes the power ratio between modes) also require ppm-level stability. 

The discrepancy between simulated and experimentally determined temperature coefficients
results from the heavy dependence on setup details such as boundary conditions at the domain
interfaces and values of material properties; assumptions made in using Eqs. (1) and (2)
could also lead to errors. Experimentally, it is difficult to isolate a single variable as can be
done with simulation, so various other destabilizing factors can play a role. However, the simulations
and experiments agree very well qualitatively, provide meaningful insight, and
with refinement could agree quantitatively as well.

\subsection{Dual-mode frequency stability}
\begin{figure}[htbp]
\centering
\includegraphics[width=0.9\linewidth]{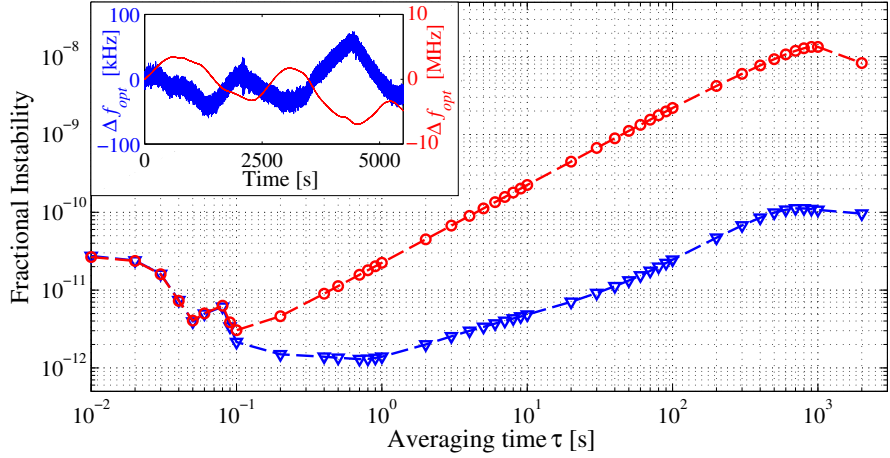}
\caption{Overlapping Allan deviation of a laser locked to the WGMR cavity, as compared to the frequency comb, under nominally identical conditions except that the dual-mode loop is closed (blue triangles) or open (red circles). Significant improvement is observed for time scales $>$ 100~ms. Inset: Time sequences of the data run showing frequency excursions for the dual-mode (blue, left scale) and free-running (red, right scale) resonator, showing periodic oscillation from the enclosure's analog temperature controller.}
\label{results}
\end{figure}

To evaluate the performance gained from the dual-mode temperature stabilization technique, we made optical comparisons to a frequency comb. After the AOM, a portion of the stabilized laser light was picked off and beat on a photodiode with a bandpass-filtered span of the comb. For several thousand seconds, the beatnote was counted with 10~ms gate times. Results of two typical data runs are shown in Fig.~\ref{results}. The lowest instability of 1.29$\times$10$^{-12}$ was reached at 700~ms, while a linear fit to the 6000 second time-series frequency excursions has a slope of 21~kHz/hr.

In Fig.~\ref{results}, we show data for both the dual-mode stabilized and ``free-running'' resonator, where the latter configuration is identical except that the temperature feedback loop is open. Stabilities below 0.1 s are identical since the dual-mode locking loop has little or no gain at the corresponding frequencies. The bump just below 0.1 s has unknown origin but is likely due to laser noises. Above 0.1 s the free running system drifts linearly with time. The dual-mode system reaches the stability floor at 1 s and then starts a random walk with time largely due to temperature gradient variation as discussed in section 3.1.

Because of these gradient effects, the effective tempco is not completely eliminated. Overall frequency stability of our dual-mode setup therefore remains limited by temperature fluctuations of the brass oven in which the resonator is mounted. An analog servo controls the oven's temperature, and the loop oscillates with a period of $\approx$2000 s. The amplitude of the oven's temperature oscillations is 2-3 mK, consistent with the frequency excursions seen in the inset of Fig.~\ref{results} for both the free running and dual-mode tempcos of -1.71 MHz/mK and 25.5 kHz/mK, respectively.  However, Fig.~\ref{results} demonstrates that for a given environmental instability, the dual-mode loop yields improvements of more than an order of magnitude within the dual-mode bandwidth, i.e. for $\tau >$1~s.

Frequency drifts caused by changes in laser power or coupling conditions remain below the limit set by environmental instabilities in our setup. Therefore, better temperature stabilization of the oven would yield better overall performance. To achieve an even higher frequency stability, temperature gradient effects must be reduced. Nevertheless, the experimentally determined effective dual-mode tempco is less than an order of magnitude larger than that of ULE cavities, and although our efforts to stabilize the environmental temperature are modest, the long-term drift rate of 21 kHz/hr is among the best stability results achieved with a whispering gallery mode reference cavity. 

\section{Conclusion}
We have investigated the frequency stability of a laser locked to a dual-mode WGMR reference cavity. Because the temperature distribution in the resonator depends on several thermal parameters simultaneously, we find that a precisely stabilized mode volume temperature does not guarantee corresponding frequency stability. Nonetheless, we calculate that the dual mode technique can reduce the temperature coefficient to 3.6$\times10^{-8}$~K$^{-1}$, comparable to the linear coefficient of ULE. In addition to environmental temperature, the level of circulating optical power also plays a significant roll. We experimentally demonstrate a reduction of temperature coefficient to 1.33$\times10^{-7}$~K$^{-1}$, a frequency instability at one second of 1.29$\times$10$^{-12}$, and a drift of 21~kHz/hr over long time periods. These results represent progress towards using WGMRs as optical references for frequency metrology applications by partially suppressing the high temperature dependence of WGMRs, thus relaxing the requirement on environmental stability. 

\section*{Acknowledgements}
This work was carried out at Jet Propulsion Laboratory, California Institute of Technology, under contract from NASA. The authors thank D. Strekalov, I.S. Grudinin, and D. Aveline for discussions and contributions to the experimental setup.

\end{document}